\documentclass[iop,apjl,preprint]{emulateapj}

\usepackage{epsf}

\newcommand{\be}{\begin{equation}}
\newcommand{\ee}{\end{equation}}
\newcommand{\bear}{\begin{eqnarray}}
\newcommand{\eear}{\end{eqnarray}}
\newcommand{\rx}{{\rm x}}

\newcommand{\rn}{{\rm n}}
\newcommand{\rp}{{\rm p}}

\newcommand{\rc}{{\rm c}}
\newcommand{\rnp}{{\rm np}}

\newcommand{\rf}{{\rm \phi}}

\newcommand{\rT}{{\rm T}}

\newcommand{\cN}{{\cal N}}
\newcommand{\cE}{{\cal E}}
\newcommand{\rP}{{\rm P}}

\newcommand{\cR}{{\cal R}}
\newcommand{\cA}{{\cal A}}
\newcommand{\cF}{{\cal F}}
\newcommand{\cK}{{\cal K}}

\shorttitle{Magneto-rotational neutron star evolution}
\shortauthors{Glampedakis \& Andersson}

\begin{document}

\lefthead{Glampedakis \& Andersson}
\righthead{Neutron star spin evolution}

\title{Magneto-rotational neutron star evolution: the role of core vortex pinning} 

\author{Kostas Glampedakis}
\affil{Theoretical Astrophysics, University of T\"ubingen, Auf der Morgenstelle 10, T\"ubingen D-72076, Germany}

\and

\author{Nils Andersson}
\affil{Mathematics, University of Southampton, Southampton SO17 1BJ, UK}
   
\begin{abstract} 
We consider the pinning of superfluid (neutron) vortices to magnetic fluxtubes
associated with a type~II (proton) superconductor in neutron star cores. We demonstrate 
that core pinning affects the spin-down of the system significantly, and
discuss implications for regular radio pulsars and magnetars. We find that
magnetars are likely to be in the pinning regime, while most radio pulsars are not.
This suggests that the currently inferred magnetic field for magnetars may be overestimated. 
We also obtain a new timescale for the magnetic field evolution which 
could be associated with the observed activity in magnetars, provided that the field has a strong 
toroidal component.

\end{abstract}



\section{Introduction} 
\label{sec:intro}

The outer core of a mature neutron star is expected to contain a 
neutron superfluid coupled to a type~II proton superconductor \citep{baym}.
The dynamics of this system is complicated; the neutrons
rotate by forming a dense array of vortices while the magnetic field is carried by quantized proton fluxtubes. 
The interaction between the vortices and the fluxtubes is expected to have significant impact on 
the dynamics of these systems. In fact, the vortices may ``pin'' to the fluxtubes
(in the sense that the energy cost of cutting through is too high) as the star spins down, affecting
the evolution of the star's rotation and magnetic field.
An example of this was provided by \citet{link03}, who argued that the  
free precession interpretation of the long-term variability in PSR B1828-11
would not be compatible with core vortex pinning.
Vortex-fluxtube pinning may also couple the magnetic field evolution to the 
rotation, as in the model of \citet{ruderman98}.

Connections between the magnetic field and the spin evolution are crucial if we want to 
understand the different observational manifestations of neutron stars, like radio pulsars
and magnetars, and establish an evolutionary link between them. 
We also need to keep in mind that the spin and its rate of change are the primary 
observables. The magnetic field tends to be inferred from 
the assumption of pure dipole breaking. 

The aim of this Letter is to demonstrate how the standard logic linking
the spin evolution and the inferred magnetic field breaks down if there is significant vortex-fluxtube pinning,
and how the same mechanism could drive the magnetic field evolution.


\section{A vortex dynamics model}

We consider the motion of individual neutron vortices and proton fluxtubes in a neutron star core, 
composed of neutrons, protons and electrons. The local vortex/fluxtube velocity is $u^i_\rx$ where the index $\rx=\{\rn,\rp \}$
distinguishes between the two types. For simplicity, the hydrodynamical 
fluid velocities are assumed to represent rigid-body rotation, i.e. $v^i_\rx = \epsilon^{ijk} \Omega_j^\rx x_k $, 
where the spin frequencies $\Omega^i_\rx$ are aligned (defining the $z$-axis) but may differ 
in magnitude. A direct consequence of uniform rotation is that the neutron vortices are straight, 
which may be a serious oversimplification. The proton fluxtubes are not constrained in this way.

In order to make progress, we make a number of simplifications. The most basic is the assumption 
of axisymmetry. 
We ignore the entrainment between the superfluids on the hydrodynamic scale
[although the effect is included implicitly, in the magnetization of neutron vortices 
\citep{als}]. Given these assumptions one can use  Amp\`ere's law, as modified in a 
type II superconductor, to show that the electron and proton fluids are ``locked'' together,
moving with a common velocity $v^i_\rp$ \citep{supercon}. Finally, we identify $\Omega_\rp$ with the motion of the ``crust'' 
and the observed spin frequency $\Omega$.

The motion of an individual vortex/fluxtube is governed by a force balance equation.
Ignoring the inertia of the filaments (as usual) we have
\bear
&&f^i_{\rm Mn} + f^i_{\rm Dn} + f^i_{\rm Pn }  =0
\label{baln}
\\
&&f^i_{\rm Mp} + f^i_{\rm Dp} + f^i_{\rm Pp} + f^i_\rT =0
\label{balp}
\eear
Each vortex/fluxtube experiences (i) a Magnus force $f^i_{\rm Mx} = \rho_\rx \epsilon^{ijk} \kappa_j^\rx ( u_k^\rx -v_k^\rx) $ 
due to the interaction between the macroscopic fluid flows and the mesoscopic, quantized,
neutron/proton circulations ($\rho_\rx$ is the fluid density while $\kappa^i_\rx $ 
has magnitude $\kappa = h/2m$ and points along the local vortex/fluxtube direction), (ii) a drag force 
$f^i_{\rm Dx} = \rho_\rx \kappa \cR_\rx ( v^i_\rp - u^i_\rx)$, typically attributed to the scattering of electrons by 
the intrinsic vortex/fluxtube  magnetic field, with a dimensionless  coefficient $\cR_\rx \ll 1$ \citep{als,ruderman98} 
and (iii) a ``pinning'' force  $f^i_{\rm Px}$ due to local interaction between 
a vortex and a fluxtube. Finally, a bent fluxtube will experience a (self-induced) tension $f^i_\rT$  
[since we are considering straight vortices there is no similar  force in (\ref{baln})]. 
All forces are per unit length. 

The direct  interaction is primarily due to the  short range magnetic fields associated with each vortex/fluxtube, 
and could lead to pinning \citep{sauls}. This interaction must obey Newton's third law, i.e.
\be
f^i_{\rm Pp} = -\alpha f^i_{\rm Pn}
\label{law3}
\ee
In fact, this relation is the operational definition of $f^i_{\rm Pp}$. In a
typical fluid element the  number density (per unit area) $\cN_\rp$ of fluxtubes greatly exceeds the vortex density $\cN_\rn$,
resulting in only a tiny fraction of fluxtubes interacting with vortices at any given instant. At the same time 
neighbouring fluxtubes can push each other,  ``distributing'' the action of the vortex 
back-reaction force $-f^i_{\rm Pn}$ to the entire fluid element. On average, this effect can be quantified by 
the small parameter  $\alpha \sim \cN_\rn/\cN_\rp$ \citep{ruderman98,jm00}.

Eqn. (\ref{baln}) provides an exact form for the pinning force,
\be
f^i_{\rm Pn} = -\rho_\rn \kappa \cR_\rn ( v^i_\rp -u^i_\rn) + \rho_\rn \kappa \epsilon^{ijk} \hat{z}_j ( v^\rn_k -u_k^\rn) 
\label{fv}
\ee
where we have set $\kappa_\rn^i = \kappa \hat{z}^i$. This can be used in (\ref{balp}) 
to eliminate $f^i_{\rm Pp}$ [with the help of eqn. (\ref{law3})]. Thus 
we obtain
\be
\tilde{\cR} ( v^i_\rp -u^i_\rn )  + \epsilon^{ijk} \hat{z}_j \left ( u_k^\rn -V_k \right ) +  \cF^i = 0  
\label{ftube2}
\ee
where we have defined
\bear
&& \tilde{\cR} \approx  \frac{\cR_\rp}{\cK},  \qquad 
 V^i = \frac{1}{\cK} \left ( \kappa_\parallel v^i_\rp
+ \frac{\alpha}{x_\rp} v^i_\rn \right ) 
\\
&& \cF^i =  \frac{1}{\cK}  \left ( \frac{1}{\rho_\rp \kappa} f^i_\rT + \cR_\rp u^i_{\rm rel} 
- \kappa_\parallel \epsilon^{ijk} \hat{z}_j u_k^{\rm rel} \right )
\eear
with  $\cK = \kappa_\parallel + \alpha/x_\rp$.
We have introduced the relative vortex-fluxtube velocity $u^i_{\rm rel} = u^i_\rn - u^i_\rp$, the ratio
$x_\rp = \rho_\rp/\rho_\rn$ (essentially the proton fraction) and the (dimensionless) projections $\kappa_\parallel$ and $\kappa_\perp$
such that
\be
\hat{\kappa}^i_\rp = \kappa_\parallel \hat{z}^i + \kappa_\perp^i, \qquad \kappa_\parallel^2 + \kappa^2_\perp = 1 
\label{k_project}
\ee
Given that the macroscopic magnetic field $B^i$ is locally directed along $\hat{\kappa}^i_\rp$, the decomposition
(\ref{k_project}) makes contact with a general ``twisted-torus'' magnetic field configuration which has 
a mixed toroidal ($B_\mathrm{T}$) and poloidal ($B_\mathrm{P}$) character
representative of hydromagnetic equilibrium  \citep{braithwaite}. 
Then $\kappa_\parallel \sim B_{\rm P}/B_{\rm T}$ and for a typical 
twisted-torus configuration $\kappa_\parallel \sim \kappa_\perp \sim 1 $. A predominantly toroidal field has
$\kappa_\parallel \ll 1 $ and $ \kappa_\perp^i \approx \hat{\varphi}^i$.
Excluding the special case $\kappa_\parallel \lesssim \alpha/x_\rp $ of a (essentially) purely toroidal field, 
we assume $\cK \approx \kappa_\parallel  $ hereafter.

As a final step, we can invert (\ref{ftube2}) and obtain the vortex velocity
\bear
&& u^i_\rn = v^i_\rp + \frac{1}{1+\tilde{\cR}^2} \Bigg [ \tilde{\cR} \epsilon^{ijk}\hat{z}_j ( v_k^\rp -V_k)
\nonumber \\
&& + \epsilon^{ijk} \epsilon_{klm} \hat{z}_j \hat{z}^l ( v^m_\rp -V^m ) 
+ \tilde{\cR} \cF^i  + \epsilon^{ijk} \hat{z}_j \cF_k \Bigg ]
\label{uvsol1}
\eear


\section{The pinning regime}
\label{sec:pin}

Strong interaction between a vortex and a fluxtube segment (via the forces $f^i_{\rm Px}$)
may result in more or less perfect pinning. Given the present understanding of neutron star matter, 
this is likely to happen \citep{sauls}. Assuming vortex pinning, in the remainder of this 
Letter we explore some very interesting consequences. If (for whatever reason) pinning is inefficient, in the sense
of a significant vortex-fluxtube relative motion, then our model does not apply.

When pinned, the two segments would move together, 
\be
u^i_\rn = u^i_\rp \equiv u^i
\ee
This velocity can be immediately obtained from (\ref{uvsol1}) after setting $u_{\rm rel}^i =0 $. 
Using cylindrical coordinates $\{\varpi,\varphi,z\}$ 
with respect to the rotation/symmetry axis, we find
\be
u^i \approx \varpi  \Omega \hat{\varphi}^i  + \frac{1}{\kappa_\parallel} \Bigg [
\varpi \frac{\alpha \Omega_\rnp}{x_\rp}  + \frac{f^\varpi_\rT}{\rho_\rp \kappa} 
\Bigg ] \left ( \frac{\cR_\rp}{\kappa_\parallel} \hat{\varpi}^i + \hat{\varphi}^i \right )
\label{upin1}
\ee
where we have defined the rotational lag $\Omega_\rnp = \Omega_\rn - \Omega$.
We see that, to leading order, the velocity of pinned vortex/fluxtube segments is nearly azimuthal and in corotation with the 
charged component. Superimposed on this there is a small relative motion $\Delta u^i \equiv u^i - v^i_\rp$ 
between the vortex/fluxtube array and the proton velocity with both radial and azimuthal components:
\bear
&&\Delta u^\varpi = u^\varpi \approx  \frac{\cR_\rp}{\kappa_\parallel^2} \left (  \varpi \frac{\alpha\Omega_\rnp}{x_\rp}  
+ \frac{f^\varpi_\rT}{\rho_\rp \kappa}  \right ) 
\label{urad1}
\\
&& \Delta u^\varphi \approx \frac{\kappa_\parallel}{\cR_\rp} u^\varpi
\label{uazim}
\eear
The dependence of $\Delta u^i$ on $\kappa_\parallel$ encodes the action of the (radial) Magnus 
force  on a fluxtube. We note that $\Delta u^i$ is nearly azimuthal unless the magnetic
field is dominated by the toroidal component to the degree that $ \kappa_\parallel \lesssim \cR_\rp \approx 10^{-3}$. 
The true magnitude of $\cR_\rp$ is  uncertain. We use a value inferred from electron scattering by a 
single fluxtube \citep{als}. 
However, we note that \citet{jones06} has argued that the high fluxtube density
suppresses the magnetic scattering leading to  a much smaller $\cR_\rp$. Nevertheless, as
no actual alternative value for $\cR_\rp$ has been provided in the literature,
we base our discussion on  the standard ``single fluxtube'' result.

A vortex cannot move arbitrarily fast and remain pinned. Above some relative vortex-fluid velocity threshold the
forces entering the balance (\ref{fv}) will exceed the maximum pinning force, $f_{\rm pin}$, and 
the vortex unpins. The maximum pinning force has been estimated to be [e.g. \citet{link03}],
\be
f_{\rm pin} \approx 3 \times 10^{15}\, B_{12}^{1/2}\, ~\mbox{dyn}/\mbox{cm}
\label{fpin}
\ee
where $B_{12} = B/10^{12}\,\mbox{G}$ is the normalized {\em interior} magnetic field.
At maximum pinning, the vortex force balance is well approximated by 
after setting $u^i_\rn \approx u^i_\rp \approx v^i_\rp$ and omitting the drag term in (\ref{fv}))
\be
f^i_{\rm Pn} \approx \rho_\rn \kappa \epsilon^{ijk} \hat{z}_j ( v^\rn_k - v^\rp_\rp ) 
\ee
and we can extract the maximum allowed spin lag (above which pinning can no longer be sustained);
\be
\Omega^{\rm max}_\rnp =  \frac{f_{\rm pin}}{\rho_\rn \kappa \varpi} \approx
7.6 \times 10^{-3} \,  \frac{B_{12}^{1/2}}{\varpi_6}\, ~\mbox{s}^{-1}
\label{maxlag} 
\ee
where $\varpi_6 = \varpi/10^6\,\mbox{cm}$.
This result can be  used in (\ref{urad1}) and (\ref{uazim}) to provide 
the maximum relative vortex/fluxtube velocity with respect to the protons;
\be
\Delta u^\varphi_{\rm max} = \frac{\kappa_\parallel}{\cR_\rp} \Delta u_{\rm max}^\varpi =  
\frac{1}{\rho_\rp \kappa \kappa_\parallel}  \left ( \alpha f_{\rm pin}  +  f^\varpi_\rT \right )  
\label{dvmax}
\ee
Let us provide numerical estimates of the forces in (\ref{dvmax}). The  tension can be
approximated by $ f_\rT = \cE_\rf/R_c$ where $\cE_\rf$ is the fluxtube energy per unit length and $R_c$ 
is the local radius of curvature. Estimates for $\cE_\rf$ can be found in \citet{mendell91} 
while a typical value for the curvature radius would be $R_c \sim R$. 
The pinning force can be estimated using (\ref{fpin}) together with $\alpha = \cN_\rn/\cN_\rp$,
$\cN_\rn = 2\Omega_\rn/\kappa $ and $\cN_\rp = B/\phi_0$ ($\phi_0 = hc/2e$ is
the flux quantum). This way we  obtain
\be
\frac{f_\rT}{\rho_\rp \kappa} \approx 10^{-9}\,
 ~\mbox{cm}/\mbox{s}, \quad
\frac{\alpha f_{\rm pin}}{\rho_\rp \kappa} \approx \frac{ 10^{-10}}{P B_{12}^{1/2}} 
\frac{\Omega_\rn}{\Omega}\,
~\mbox{cm}/\mbox{s}
\label{fTvel}
\ee
with the stellar spin period, $P$, measured in seconds.


\section{Spin evolution}
\label{sec:spin}

Vortex pinning has direct impact on the neutron star spin evolution. 
This is not surprising, given the close link between the spindown of the neutron superfluid 
and the radial vortex motion,
\be
u^\varpi_\rn = -\varpi {\dot{\Omega}_\rn}/{2\Omega_\rn}
\label{unrad}
\ee
If the neutron superfluid is able to track the electromagnetically-driven spindown of the charged component 
(i.e. when $\Omega_\rn = \Omega $ and $\dot{\Omega}_\rn = \dot{\Omega}$), we have 
\be
u_\rn^\varpi \approx 8 \times 10^{-7} \varpi_6 
\left ( {10^4\,\mbox{yr}}/{\tau_{\rm sd}} \right )\, \mbox{cm}/\mbox{s}
\ee 
where $\tau_{\rm sd} = P/2|\dot{P}|$ is the observed ``spin-down age''. This vastly exceeds 
the maximum radial vortex velocity  (\ref{dvmax}) in the pinning regime.
Thus, we should expect  the neutrons to spin down at a rate $|\dot{\Omega}_\rn| \ll |\dot{\Omega}|$ 
in the pinning regime. This would lead to a monotonically increasing spin lag $\Omega_\rnp$, 
to the point where $\Omega_\rnp^{\rm max}$ is reached and vortex pinning can no longer be sustained. 
The development of this lag is similar in most models of large pulsar glitches.

Let us focus on the timescale associated with the pinning phase. In order to
estimate this, we  need to model the spin evolution of a multi-fluid star. 
This can be done by considering the volume-integrated Euler equations for the fluids 
[cf. \citet{trev}].
The resulting equations take the form,
\be
I_\rn \dot{\Omega}_\rn = N_{\rm mf}, \qquad I_\rp \dot{\Omega} = -N_{\rm mf} + N_{\rm em}
\label{spinevol}
\ee
where $I_\rx = \int  \rho_\rx \varpi^2 dV$ is the moment of inertia for each fluid. 
These equations feature torques due to (i) the exterior magnetic field ($N_{\rm em}^i$) and (ii) the ``mutual friction''
($N_{\rm mf}^i$) which represents the vortex/fluxtube-mediated coupling. The latter torque is given by  
\be
N^i_{\rm mf}  = \int \epsilon^{ijk} x_j F_k^{\rm mf}  dV
\label{torque}
\ee
where the mutual friction force is
\be
 F^i_{\rm mf} = \cN_\rn \left ( f^i_{\rm Dn} + f^i_{\rm Pn} \right ) =
2 \Omega_\rn \rho_\rn \epsilon^{ijk} \hat{z}_j ( v^\rn_k -u_k ) 
\ee
Using our result (\ref{upin1}) for the common vortex/fluxtube velocity in the pinning regime
we have
\be
N^i_{\rm mf} \approx -2\Omega_\rn \hat{z}^i   \int \varpi^2 \frac{\rho_\rn \cR_\rp}{\kappa^2_\parallel} 
\left (  \frac{\alpha \Omega_\rnp}{x_\rp} + \frac{f^\varpi_\rT}{\varpi \rho_\rp \kappa} \right )  dV  
\label{Nmf}
\ee

We also need to consider the magnetic torque $N^i_{\rm em}$. Following the analysis of 
\citet{spit}, which accounts for a wind contribution, we have
\be
N_{\rm em}^i = \left ( \frac{B_d^2 R^6}{4c^3} \right ) \Omega^3 \hat{z}^i  
\equiv \cA \Omega^3 \hat{z}^i 
\label{Nem}
\ee
where $B_d$ is the {\em exterior dipole} field at the magnetic pole. 

Returning to the spin evolution equations (\ref{spinevol}) with (\ref{Nmf}) and (\ref{Nem}), we observe
that the timescale $\tau_{\rm mf} \sim I_\rn \Omega /N_{\rm mf}$ associated with the pinning and tension forces 
is much longer than the electromagnetic spindown timescale. For typical neutron star parameters
we find $ \tau_{\rm mf} \sim 10^{10}\, \kappa^2_\parallel\, \mbox{yr}$.
Hence, the spin evolution  in the perfect pinning regime is 
well approximated by
\be
\dot{\Omega}_\rn \approx 0, \qquad \dot{\Omega} \approx -{\cA} \Omega^3 / {I_\rp}
\label{spinevol1}
\ee
in accordance with the intuitive notion of a limited radial vortex motion enforced by pinning,
and of the magnetic braking affecting only the charged component.
Solving (\ref{spinevol1}) with initial data $\Omega_\rx (0)$ is straightforward, and we have
\be
\Omega (t) = \Omega(0) \Bigg ( 1 +  \frac{I_0}{I_\rp}\frac{t}{\tau_0}\Bigg )^{-1/2}, \qquad \Omega_\rn = \Omega_\rn(0)
\label{spinsol}
\ee
where $I_0 = I_\rn + I_\rp$ is the total moment of inertia and $\tau_0 = I_0/(2\cA \Omega^2) $.
In a  single-component star, $\tau_0$ would correspond to the spin-down 
timescale and would be identified with the observed spin-down age $\tau_{\rm sd}$.
However, in the superfluid pinning model the appropriate identification, based on the spindown law (\ref{spinevol1}), is 
$\tau_{\rm sd} =  (I_\rp/I_0) \tau_0 \approx x_\rp \tau_0$. Equivalently, we can relate the {\em true} field $B_d$ with the
field $B_d^{\rm inf}$ {\em inferred} from spin-down,
\be
B_d = \left (\frac{I_\rp}{I_0} \right )^{1/2} \left ( \frac{c^{3} I_0}{\pi^2 R^6} |\dot{P}| P \right )^{1/2} 
\approx x_\rp^{1/2} B_d^{\rm inf}
\label{age}
\ee
The spin evolution (\ref{spinsol})  naturally leads to a monotonically increasing 
spin-lag. Considering, for instance, the early  regime $ t \ll (I_\rp/I_0) \tau_0  $, we can easily
obtain the time $t_{\rm max}$ at which the system reaches the maximum lag  (\ref{maxlag}), where vortices first unpin
[assuming  $\Omega_\rnp (0) =0$]
\be
\frac{t_{\rm max}}{\tau_{\rm sd}} \approx \frac{2\Omega_\rnp^{\rm max}}{\Omega} 
\approx 2.4 \times 10^{-3}\,\frac{P  B_{12}^{1/2}}{\varpi_6} 
\label{tmax1}
\ee
which features the interior $B$ field.
When applied to a typical radio pulsar ($P=0.1\,\mbox{s}$, $B =10^{12}\,\mbox{G}$) the result (\ref{tmax1}) 
leads to  $t_{\rm max} \ll \tau_{\rm sd}$. 
Thus, core vortex pinning in a typical pulsar is an ephemeral phenomenon,
plausibly  associated with large glitches \citep{glitchlett}. 

The situation may be very different for magnetars. For a strongly magnetized and slowly spinning
object the timescale $t_{\rm max}$ can be comparable to, or even exceed, $\tau_{\rm sd}$. In this case
we need to resort to the full solution (\ref{spinsol}). Inserting canonical magnetar parameters 
($B = 10^{15}\,\mbox{G}, P = 10\,\mbox{s}$), we obtain 
\be
t_{\rm max} \gtrsim \tau_{\rm sd}
\label{tmax2}
\ee
This suggests that  vortex-fluxtube pinning may well {\em persist} in magnetar cores.


\section{Implications and discussion}
\label{sec:dac}

We have provided a rather simplistic analysis of a very complex problem, Yet, the results
could have important implications for the magneto-rotational properties of neutron stars. 
Perhaps the most significant conclusion concerns the potential persistence of vortex pinning in 
magnetars, as suggested by (\ref{tmax2}). This implies (see eqn.~(\ref{age})) that the surface dipole field inferred from spin-down  
is an {\em overestimate} by a factor  $ (I_\rp/I_0)^{-1/2} \approx x_\rp^{-1/2} \approx 3 - 10$. 
This  mechanism could be responsible
for the apparent clustering of the observed magnetar spin periods, since the end of the pinning regime 
would be accompanied by an effective weakening of the torque; $N_{\rm em} \to x_\rp N_{\rm em}$. 
Interestingly,  the presence of 
a ``twisted'' magnetar magnetosphere may lead to a similar effect \citep{TLK02}.

There are also implications for the magnetic field evolution in magnetar cores.
According to the pinning model the motion of the proton fluxtubes (and therefore of the magnetic field)
is controlled by the neutron vortices and vice versa. We can define a magnetic evolution timescale
associated with the relative vortex/fluxtube motion with respect to the proton fluid,
$\tau_B = {L}/|\Delta u_{\rm max}|$, where $L$ is some typical lengthscale. In magnetars $\Delta u_{\rm max}$ is mostly due to 
the tension since (\ref{fTvel}) implies $ f_\rT \gg \alpha f_{\rm pin} $ [note that
(\ref{maxlag}) suggests $\Omega_\rn \sim \Omega$ for magnetars].
The shortest timescale is the one associated with the azimuthal motion
$\Delta u^\varphi_{\rm max}$, leading to
\be
\tau_B \sim \frac{\kappa_\parallel \kappa \rho_\rp L}{f_\rT} \sim 10^6\, L_5 
\frac{B_\rP}{B_\mathrm{T}} \left (\frac{R_\rc}{10\,\mbox{km}} \right )\,\mbox{yr}
\label{tauB1}
\ee
where we have restored the explicit dependence of $f_\rT$ on the curvature radius $R_\rc$ and defined $L_5 = L/10^5\mbox{cm}$. 
The timescale associated with the radial fluxtube motion is longer by a factor $\cR_\rp^{-1} \sim 10^3$. 
Also, $\tau_B$ becomes significantly longer if (for whatever reason) $f_\rT$ is negligible.

Is the timescale in (\ref{tauB1}) relevant for the observed 
magnetar population? For this to be the case we need $\tau_B$ to be comparable to the age 
$\tau_{\rm sd} \sim 10^3-10^4\,\mbox{yr}$ of these objects. At first sight,
 (\ref{tauB1}) seems to suggest a  significantly longer timescale. 
However, our result approaches the  interesting regime if we consider 
a magnetic field with a locally strong toroidal component $B_\mathrm{T} >  B_\rP$ and a curvature $R_\rc \lesssim 0.1 R$. 
Indeed, recent work on stable hydromagnetic equilibria \citep{braithwaite} suggests that these properties could 
be realistic.

Our results indicate that in the regime of persistent vortex pinning in magnetar cores the evolution 
of a strong toroidal magnetic field component
is {\em accelerated} with respect to 
that of the poloidal component, and could occur on a timescale $10^4-10^5\,\mbox{yr}$.
Comparing to recent work on ambipolar diffusion in superfluid neutron stars 
\citep{ambipol}, we see that persistent vortex pinning is the main driving force of magnetic field evolution 
in magnetar cores, with possible relevance to the activity observed in these objects. 
We also note that $\tau_B$ can be comparable to the ambipolar diffusion timescale in weakly/non-superfluid
magnetars (e.g. \citet{act04}). 

The implications for  radio pulsars are more subtle. Recalling
 equation (\ref{tmax1}), core vortex pinning may only be supported over short periods of a pulsar's life. 
Thus, pulsars likely  spend most of their time in a ``cutting regime'' where vortices can 
move through the fluxtubes \citep{link03}. 
This result is in apparent conflict with the work by 
\citet{ruderman98} and \citet{jm00}. In these studies it is  assumed that  the radial 
vortex  motion is given by (\ref{unrad}) with $\dot{\Omega}_\rn = \dot{\Omega}$ and $\Omega_\rn = \Omega$ in
the pinning regime. Our discussion shows that this assumption is not valid. Moreover, the previous studies do not 
take full account of the Magnus force. Hence, the results of \citet{ruderman98} and \citet{jm00}
must be considered with caution.

In this Letter we have discussed issues that are central to our understanding of the 
magneto-rotational evolution of neutron stars. We have shown that the observed magnetars may be in the regime 
of core vortex pinning, affecting the bulk spin-down, the inferred surface magnetic field and the magnetic field evolution. 
Our results also show that core pinning is likely less relevant in 
the weaker magnetic field radio pulsars.

The issues we have discussed obviously need more detailed consideration. 
Future work needs to relax many of our assumptions. First of all, we have assumed that the entire star 
is in a superfluid state while, in reality, this may be an accurate description only for the outer core 
 -- in that case $B_d$ would be overestimated by the smaller factor $\sim \left ( (I_\rp + I_{\rm N})/I_0 \right )^{-1/2}$ 
where $I_{\rm N}$ is the moment of inertia of the non-superfluid neutrons. 
To make progress we need to consider a model with realistic stratification. 
For the magnetars in particular, it not not yet clear that  
their core temperature  is below the level for the onset of neutron superfluidity
suggested by the recent cooling models for the Cassiopeia A neutron star \citep{page11,shtern11}.
The most detailed work on magnetar temperature profiles \citep{kam06} appears to favour superfluidity, 
but the issue is far from settled \citep{act04,dss09}.
A related case for which our model does not apply is for a bulk $B$ field in excess of 
the critical value $\approx 10^{16}\,\mbox{G}$ above which proton superconductivity is suppressed \citep{baym}. 
We also need to remove the assumption of uniform rotation, which does not remain
consistent  when the long-term neutron star evolution is considered.
This would help link our vortex/fluxtube model to the physics
of ambipolar diffusion \citep{ambipol}.
The departure from uniform rotation is also likely to be important in the vortex 
cutting regime, which should be relevant for normal pulsars.
These issues require us to consider more realistic neutron star models,  a challenge that should ultimately reward 
us with a deeper understanding of these exciting systems.
\acknowledgements
KG is supported by an Alexander von Humboldt fellowship and by the German DFG via SFB/TR7.
NA acknowledges support from STFC in the UK. 


\end{document}